\documentclass[10pt, conference, compsocconf]{IEEEtran}
\usepackage[american]{babel}
\usepackage[latin1]{inputenc}
\usepackage{amsmath,amsfonts,amssymb}
\usepackage{paralist}
\usepackage{url,xspace}
\usepackage[ruled,vlined]{algorithm2e}
\usepackage{graphicx,color,psfrag}
\graphicspath{{fig/}}
\usepackage{pstool}
\usepackage{caption}[2007/12/23]
\DeclareCaptionType{copyrightbox}
\usepackage{multirow}
\usepackage{ifthen}
\usepackage{subfigure}

\newcommand{\elim}{\mathit{elim}}

\newcommand{\GEQRT}{\ensuremath{\mathit{GEQRT}}\xspace}
\newcommand{\TSQRT}{\ensuremath{\mathit{TSQRT}}\xspace}

\newcommand{\UNMQR}{\ensuremath{\mathit{UNMQR}}\xspace}
\newcommand{\TSMQR}{\ensuremath{\mathit{TSMQR}}\xspace}
\newcommand{\TTQRT}{\ensuremath{\mathit{TTQRT}}\xspace}
\newcommand{\TTMQR}{\ensuremath{\mathit{TTMQR}}\xspace}

\newcommand{\s}{\ensuremath{\star}\xspace}

\newcommand{\SK}{\textsc{Sameh-Kuck}\xspace}
\newcommand{\MC}{\textsc{Fibonacci}\xspace}
\newcommand{\BT}{\textsc{BinaryTree}\xspace}
\newcommand{\Greedy}{\textsc{Greedy}\xspace}

\newcommand{\FT}{\textsc{FlatTree}\xspace}

\newcommand{\killer}{\ensuremath{\mathit{killer}}\xspace}
\newcommand{\TS}{\ensuremath{\mathit{TS}}\xspace}
\newcommand{\TT}{\ensuremath{\mathit{TT}}\xspace}
\newcommand{\CYCLIC}{\ensuremath{\mathit{CYCLIC}}\xspace}
\newcommand{\Dague}{\textsc{DAGuE}\xspace}

\newcommand{\scalapack}{{\sc ScaLAPACK}\xspace}
\newcommand{\blas}{{\sc BLAS}\xspace}

\newcommand{\hqr}{{\sc HQR}\xspace}

\usepackage{wasysym}

\def \longversion{0}

\newcommand{\OnlyRR}[1]{ \ifnum\longversion=1 #1  \fi }
\newcommand{\OnlyIPDPS}[1]{ \ifnum\longversion=0 #1  \fi }

\newboolean{withproof}
\setboolean{withproof}{false}
\newboolean{withoutproof}
\setboolean{withoutproof}{true}
\newcommand{\forRR}[1]{\ifthenelse{\boolean{withproof}}{{#1}}{}}
\newcommand{\forSCpaper}[1]{\ifthenelse{\boolean{withoutproof}}{{#1}}{}}

\title{Hierarchical QR factorization algorithms\\
for multi-core cluster systems}

\author{
\IEEEauthorblockN{Jack Dongarra$^{1,2,3}$, Mathieu Faverge$^1$, Thomas Herault$^1$, Julien Langou$^4$ 
and Yves Robert$^{1,5}$}
\IEEEauthorblockA{1. University of Tennessee Knoxville, USA\\
2. Oak Ridge National Laboratory, USA\\
3. Manchester University, UK\\
4. University of Colorado Denver, USA; supported by the National Science Foundation grant \# NSF CCF 1054864.  \\
5. Ecole Normale Sup\'erieure de Lyon, France\\
\{dongarra | mfaverge | therault\}@eecs.utk.edu, Julien.Langou@ucdenver.edu, Yves.Robert@ens-lyon.fr
}
}

\begin{document}

\maketitle

\begin{abstract}
%
%
\OnlyRR{
This paper describes a new QR factorization algorithm which is especially
designed for massively parallel platforms combining parallel distributed
multi-core nodes. 
These platforms make the
present and the foreseeable future of high-performance computing. 
 Our new QR
factorization algorithm falls in the category of the tile algorithms which
naturally enables good data locality for the sequential kernels executed by the
cores (high sequential performance), low number of messages in a parallel
distributed setting (small latency term), and fine granularity (high
parallelism).  Each tile algorithm is uniquely characterized by its sequence of
reduction trees. In the context of a cluster of multicores, in order to
minimize the number of inter-processor communications
(aka, ``communication-avoiding'' algorithm), it is natural to consider two-level hierarchical
trees composed of an ``inter-node'' tree which acts on top of ``intra-node''
trees. At the intra-node level, we propose a hierarchical tree made of three
levels: (0) ``TS level'' for cache-friendliness, (1) ``low level'' for decoupled
highly parallel inter-node reductions, (2) ``coupling level'' to efficiently
resolve interactions between local reductions and global reductions.
Our hierarchical algorithm
and its implementation are flexible and modular, and can accommodate several
kernel types, different distribution layouts, and a variety of reduction trees
at all levels, both inter-cluster and intra-cluster.
Numerical experiments on
a cluster of multicore nodes (1) confirm that each of the four levels of our hierarchical tree contributes to
build up performance and (2) build insights on how these levels influence performance
and interact within each other.  Our implementation of the new algorithm with the \Dague
scheduling tool significantly outperforms currently available QR factorization
softwares for all matrix shapes, thereby bringing a new advance in numerical linear algebra for
petascale and exascale platforms. We compare to the \scalapack library and to two parallel distributed QR factorization
for multicore clusters, both published less than a year ago.
On tall and skinny matrices, we reach 57.5\% of theoretical computational peak performance
to be compared with 
6.4\% for \scalapack (9.0x speedup), 
18.3\% for [BDD+10] (3.1x), 
43.5\% for [SLHD10] (1.3x).
On square matrices, we reach 68.7\% of theoretical computational peak performance
to be compared with 
44.2\% for \scalapack (1.6x), 
62.2\% for [BDD+10] (1.1x), 
46.7\% for [SLHD10] (1.5x).

}
\OnlyIPDPS{
This paper describes a new QR factorization algorithm which is especially
designed for massively parallel platforms combining parallel distributed
multi-core nodes. 
These platforms represent the
present and the foreseeable future of high-performance computing. 
 Our new QR
factorization algorithm falls in the category of the tile algorithms which
naturally enables good data locality for the sequential kernels executed by the
cores (high sequential performance), low number of messages in a parallel
distributed setting (small latency term), and fine granularity (high
parallelism).  Each tile algorithm is uniquely characterized by its sequence of
reduction trees. In the context of a cluster of multicores, in order to
minimize the number of inter-processor communications
(aka, ``communication-avoiding''), it is natural to consider hierarchical
trees composed of an ``inter-cluster'' tree which acts on top of ``intra-cluster''
trees. At the intra-cluster level, we propose a hierarchical tree made of three
levels: (0) ``TS level'' for cache-friendliness, (1) ``low level'' for decoupled
highly parallel inter-node reductions, (2) ``coupling level'' to efficiently
resolve interactions between local reductions and global reductions.
Our hierarchical algorithm
and its implementation are flexible and modular, and can accommodate several
kernel types, different distribution layouts, and a variety of reduction trees
at all levels, both inter-cluster and intra-cluster.
Numerical experiments on
a cluster of multicore nodes (i) confirm that each of the four levels  of our hierarchical tree contributes to
build up performance and (ii) build insights on how these levels influence performance
and interact within each other.  Our implementation of the new algorithm with the \Dague
scheduling tool significantly outperforms currently available QR factorization
softwares for all matrix shapes, thereby bringing a new advance in numerical linear algebra for
petascale and exascale platforms. 

}

\end{abstract}
\begin{IEEEkeywords}
QR factorization; numerical linear algebra; hierarchical architecture; distributed memory; cluster; 
multicore.
\end{IEEEkeywords}

\section{Introduction}
\label{sec.intro}

Future exascale machines will likely be massively parallel architectures, with
$10^5$ to $10^6$ processors, each processor itself being equipped with $10^3$
to $10^4$ cores
 At the node level, the
architecture is a shared-memory machine, running many parallel threads on the
cores.
At the machine level, the architecture is a
distributed-memory machine.  This additional level of hierarchy, together with
the interplay between the cores and the accelerators, dramatically complicates
the design of new versions of the standard factorization algorithms that are
central to many scientific applications.  In particular, the performance of
numerical linear algebra kernels is at the heart of many grand challenge
applications, and it is of key importance to provide highly-efficient
implementations of these kernels to leverage the impact of exascale platforms.

This paper investigates the impact of this hierarchical hardware
organization on the design of numerical linear algebra algorithms. We deal with
the QR factorization algorithm which is ubiquitous in high-performance
computing applications, and which is representative of many numerical linear
algebra kernels.  In recent years, the quest of efficient, yet portable,
implementations of the QR factorization algorithm has never
stopped~\cite{sc-paper2011,SongLHD10,tsqr-grid,Buttari2008,Quintana:2009,CAQR,Hadri_ipdps_2010,dague-la}.
In a nutshell, state-of-the-art software has evolved from block-column panels
to tile-based versions, and then to multi-killer algorithms.  We briefly review
this evolution in the following paragraphs. 

First the LAPACK library~\cite{lawn41} has provided  Level~3 \blas kernels to
boost performance on a single CPU. The \scalapack library~\cite{Scalapack97}
builds upon LAPACK and provides a coarse-grain parallel version, where
processors operate on large block-column panels, i.e.  blocks of $b$ columns of
the original matrix. Here $b$ is the block size, typically $b=200$ or more, for
Level~3 BLAS performance.  Inter-processor communications occur through highly
tuned MPI send and receive primitives.  The factorization progresses panel by
panel.  Once the current panel is factored, updates are applied on all the
following panels (remember that the matrix operated upon shrinks as the
factorization progresses).  Sophisticated \emph{lookahead} versions of the
algorithms factor the next panel while the current updates are still being
applied to the trailing matrix.

Then, the advent of multi-core processors has led to a major modification in the
algorithms~\cite{Buttari2008,Quintana:2009,Hadri_ipdps_2010,sc-paper2011}.  Now
each processor should run several threads in parallel to keep all cores within
that processor busy.  Tiled versions of the algorithms have thus been designed:
dividing large block-column panels into several tiles allows for a decrease in the
granularity down to a level where many smaller-size tasks are spawned.  In the
current panel, the diagonal tile, or killer tile, is used to kill all the tiles
below it in the panel.  Because the factorization of the whole panel is now
broken into the killing of several tiles, the update operations can also be
partitioned at the tile level, which generates many tasks to feed all cores.
However, the dependencies between these  tasks must be enforced, and the
algorithms have become much more complicated. 

A technical difficulty arises with the killing operations within the panel:
these are sequential because the diagonal tile is used for each of them, hence
it is modified at each killing operation.  This observation applies to the
updates as well: in any trailing column, the update of a tile must wait until
the update of its predecessor tile is completed. To further increase the degree
of parallelism of the algorithms, it is possible to use several killer tiles
inside a panel. The only condition is that concurrent killing operations must
involve disjoint tile pairs.  Of course, in the end there must remain only one
non-zero tile on the panel diagonal, so that all killers but the diagonal tile
must be killed themselves later on, using a reduction tree of arbitrary shape
(e.g. serial, fully binary, ...).  The extra source for
parallelism resides in the fact that the whole matrix can be partitioned into
domains, with one killer per domain responsible for killing the tiles local to
the domain.  In each domain, all these operations, and the corresponding
updates, are independent and can run concurrently.  Such multi-killer
algorithms represent the state-of-the-art for multi-core processors, but they
are still being refined, because the impact of the reduction trees which are
chosen is not fully understood, and also because using many killers implies the
use of different tile kernels, called TT kernels, which are less-efficient than
the TS-kernels used with a single killer per panel.

The goal of this paper is to move a step forward and to introduce a
flexible and modular algorithm for clusters of multi-core processors. Tackling
such hierarchical architectures is a difficult challenge for two
reasons. The first challenge arises from the algorithmic perspective. Brand
new avenues must be explored to accommodate the hierarchical nature of
multi-core cluster systems.  Concurrent killers allow for more parallelism, but
the reduction tree that follows breaks the smooth pipelining of operations from
one panel to the next. With one domain per processor, we may have not enough
parallel operations to feed all the many-cores, so we may need to have several
domains per processor.  The reduction operations involve inter-processor
communications, which are much slower than intra-processor shared memory
accesses. Limiting their number could be achieved with a block
row-distribution, but this would severely imbalance processor loads. This small
list is not exhaustive: good load-balance, efficient pipelining, and memory
locality are all conflicting objectives. The main contribution of this paper is
to provide a novel algorithm that is fully aware of the hierarchical nature of
the target platform and squeezes the most out of its resources.

The second challenge is at the programming level.  Within a multi-core
processor, the architecture is a shared-memory machine, running many parallel
threads on the cores.  But the global architecture is
a distributed-memory machine, and requires MPI communication primitives for
inter-processor communications.  A slight change in the algorithm, or in the
matrix layout across the processors, might call for a time-consuming and
error-prone process of code adaptation. For each version, one must  identify,
and adequately implement, inter-processor versus intra-processor kernels. This
dramatically complicates the task of the programmer if she relies on a manual
approach. We solve this problem by relying on the \Dague software~\cite{dague-engine, dague-la}, 
so that we
can concentrate on the algorithm and forget about MPI sends and thread
synchronizations. Once we have specified the algorithm at a task level, the
\Dague software will recognize which operations are local to a processor (and
hence correspond to shared-memory accesses), and which are not (and hence must
be converted into MPI communications).  Our experiments show that this approach
is very powerful, and that the use of a higher-level framework does not prevent
our algorithms from outperforming all existing solutions.

In this paragraph, we briefly highlights our contribution with respect to existing work
(see Section~\ref{sec.relatedwork} for a full overview). 
Two recent papers \cite{dague-la,SongLHD10} have discussed tiled
algorithms for clusters of multicore. In~\cite{SongLHD10}, the authors use a
two-level hierarchical tree made of an inter-node binary tree on top of an
intra-node TS flat tree and use a 1D block data layout. The limitations are: (1)
the use of a flat tree at the node level is not adapted when the local matrices
are tall and skinny; (2) the use of the 1D block data layout results in serious
load imbalance for square matrices. In~\cite{dague-la}, the authors use a plain
flat tree on top of a 2D block data layout. The limitations are: (1) the use of
a flat tree is not adapted for tall and skinny matrices; (2) the flat tree with natural
ordering is not aware of the 2D data block cyclic distribution and therefore
performs many more communications than needed. Our algorithm addresses all these
issues while keeping the positive features.  At the intra-node level, we
propose a hierarchical tree made of three levels: (0) ``TS level'' for
cache-friendliness, (1) ``low level'' for decoupled highly parallel inter-node
reductions, (2) ``coupling level'' to efficiently resolve interactions between
local reductions and global reductions. Finally (3) a ``high level'' tree is
used for the inter-node reduction. The use of the ``high level'' tree enables a
small number of interprocessor communications, thereby making our algorithm
``communication-avoiding''.  For the levels (1), (2) and (3) of the hierarchical
algorithm, the reduction can accommodate any tree.  Our implementation is
flexible and modular, and proposes several reduction trees per level.  This
allows us to use those reduction trees which are efficient for a given matrix shape. Finally
the ``coupling level'' -- which operates within a node, and fits in between the
intra- and inter-cluster reductions -- resolves all interactions between
the low level and  high level trees,  in such a way that the low level tree
(acting within a cluster) becomes decoupled from the influence of the other clusters.  
To summarize, our new algorithm is a tile QR factorization which is (a) designed especially for  massively parallel platforms combining parallel distributed
multi-core nodes; (b) features a hierarchical four-level tree reduction; 
(c) incorporates a novel coupling level; (d) is 2D block cyclic aware; and (e) implements 
a variety of trees at each level. The resulting properties of the algorithm are
(i) cache-efficiency at the core level, 
(ii) high granularity at the node level,
(iii) communication avoiding at the distributed level, 
(iv) excellent load balancing overall,
(v) nice coupling between the inter-node and intra-node interactions,
and (vi) ability to efficiently handle any matrix shape.

The rest of the paper is organized as follows. We start with a quick review of
tiled QR algorithms (Section~\ref{sec.review}). Then we  detail the key principles underlying the
design of state-of-the-art algorithms from the literature (Section~\ref{sec.relatedwork}).
 The core
contributions of the paper reside  in Section~\ref{sec.newalgo}, where we
describe our new algorithm in full details, and  in Section~\ref{sec.expe},
where we provide experiments showing that we outperform current
state-of-the-art implementations.
 
\begin{algorithm}[!htb]
  \DontPrintSemicolon
  \For{$\textnormal{k} = 0$ to $\min(m,n)-1$}{
     \For{$\textnormal{i} = k+1$ to $m-1$}{
     $\elim(i, \killer(i,k), k)$\;
    }
  }
\caption{Generic QR algorithm.}
\label{alg.QR}
\end{algorithm}

\begin{algorithm}[!htb]
  \DontPrintSemicolon
  $\GEQRT(killer(i,k), k)$\\
  $\TSQRT(i,killer(i,k), k)$\\
  \For{$\textnormal{j} = k+1$ to $n-1$}{
     $\UNMQR(killer(i,k), k, j)$\\
     $\TSMQR(i, killer(i,k), k, j)$
  }
  (a) \emph{With TS (Triangle on top of square)} kernels\\
  ~\\
    $\GEQRT(killer(i,k), k)$\\
  $\GEQRT(i, k)$\\
  \For{$\textnormal{j} = k+1$ to $n-1$}{
     $\UNMQR(killer(i,k), k, j)$\\
     $\UNMQR(i, k, j)$
  }
  $\TTQRT(i, killer(i,k), k)$\\
  \For{$\textnormal{j} = k+1$ to $n-1$}{
     $\TTMQR(i, killer(i,k), k, j)$\\
     }
   (b) With \emph{TT (Triangle on top of triangle)} kernels.
\caption{Elimination $\elim(i, killer(i,k), k)$.}
\label{alg.elimSQTR}
\end{algorithm}

\section{Tiled QR algorithms}
\label{sec.review}

The general shape of a tiled QR algorithm for a tiled matrix of $m \times n$ tiles, whose rows and columns are indexed
from $0$, is given in Algorithm~\ref{alg.QR}. Here $i$ and $k$ are tile indices, and we have square $b \times b$ tiles,
where $b$ is the block size. Thus the actual size of the matrix is  
$M \times N$, with $M=m*b$ and $N=n*b$. The first loop index 
$k$ is the panel index, and $\elim(i, \killer(i,k), k)$
is an orthogonal transformation that combines rows $i$ and $\killer(i,k)$ to zero
out the tile in position $(i,k)$. Each elimination $\elim(i, \killer(i,k), k)$
consists of two substeps: (i) first in column $k$, tile $(i,k)$ is zeroed out (or killed) by tile
$(\killer(i,k),k)$; and (ii) in each following column $j>k$, tiles $(i,j)$ and
$(\killer(i,k),j)$ are updated; all these updates are independent and can be triggered as soon as the
killing is completed.
The algorithm is entirely characterized by its
\emph{elimination list}, which is the ordered list of all the eliminations
$\elim(i, \killer(i,k), k)$ that are executed. 

To implement an orthogonal transformation $\elim(i,killer(i,k), k)$, we can use either \TT kernels or \TS kernels,
as shown in Algorithm~\ref{alg.elimSQTR}.
In a nutshell, a tile can have three states: square, triangle,
and zero. Initially, all tiles are square. A killer must be a triangle, and we transform a square into a triangle
using the \GEQRT kernel. 
With a single killer, we start by transforming it into a triangle (kernel \GEQRT) before eliminating square tiles.
To kill a square with a triangle, we use the \TSQRT kernel.
With several killers, we have several triangles, hence the need for an additional kernel
 to eliminate a triangle (rather than a square): this
is the \TTQRT kernel.
The number of arithmetic operations performed by a \TSQRT kernel (to kill a square)
is the same as that of a \GEQRT
(transform the square into a triangle) followed by a \TTQRT (kill a triangle).
The same observations basically applies for the corresponding updates,
which can be decomposed in a similar way (see Algorithm~\ref{alg.elimSQTR}).
The \TS kernels can only be used within a flat tree at the first tree level (so that tiles are square).
On the one hand, \TT kernels offer more
parallelism than \TS kernels. On the other hand, the sequential performance of the \TS kernels is higher
(e.g., by 10\% in our experimental section) than the one of the \TT kernels.
We refer to~\cite{sc-paper2011} for more information on the various kernels.

Any tiled QR algorithm used to factor a tiled matrix
of $m \times n$ tiles is characterized by its elimination list.
Obviously, the algorithm must zero out all tiles below the diagonal: for each tile $(i,k)$, $i>k$,
$0 \leq k < \min(m,n)$, the list must contain exactly one entry $\elim(i, \s, k)$, where \s denotes some
row index $killer(i,k)$ . There are two conditions for a transformation $\elim(i, killer(i,k), k)$ to be valid:
$\bullet$ both rows $i$ and $killer(i,k)$ must be ready, meaning that all their tiles
      left of the panel (of indices $(i,k')$ and $(killer(i,k),k')$ for $0 \leq k' < k$)
      must have already been zeroed out: all transformations $\elim(i, killer(i,k'), k')$
      \linebreak and $\elim(killer(i,k), killer(killer(i,k),k'), k')$ must precede
      $\elim(i, killer(i,k), k)$ in the elimination list
$\bullet$ row $killer(i,k)$ must be a potential annihilator, meaning that tile
      $(killer(i,k),k)$ has not been zeroed out yet: the transformation
      $\elim(killer(i,k), killer(killer(i,k),k), k)$ must follow $\elim(i, killer(i,k), k)$ in the elimination list.

Assuming square $b$-by-$b$ tiles and using a $b^3/3$ floating point operation unit,
the weight of \GEQRT is 4, \UNMQR 6, \TSQRT 6, TSMQR 12, \TTQRT 2, and \TTMQR 6.
A critical result is that no matter what elimination list is used, or which kernels are called, the total
weight of the tasks for performing a tiled QR factorization algorithm is
constant and equal to $6 m n^2 - 2 n^3$. Using 
$M = m * b$, and $N = n * b$, we retrieve $2 M N^2 - 2/3 N^3 $ floating point operations, the
exact same number as for a standard Householder reflection algorithm as found
in LAPACK (e.g.,~\cite{lawn41}).
In essence, the execution of a  tiled QR algorithm is fully determined by
its elimination list.  
Each
transformation involves several kernels, whose execution can start as soon as they
are ready, i.e., as soon as all dependencies have been enforced.
\section{Related work}
\label{sec.relatedwork}

In this section, we survey tiled QR algorithms from the literature, and we outline their main characteristics. 
We start with several examples to help the reader better understand the combinatorial
space that can be explored to design such algorithms.

\subsection{Factoring the first panel}

In this section we discuss several strategies for factoring the first panel, of index $0$, of 
a tiled matrix of $m \times n$ tiles. When designing an efficient
algorithm, 
individual panel factorization should not be considered separately from the rest of the factorization, but concentrating on
a single panel is enough to illustrate several important points.

\begin{figure}[htb]
\includegraphics[width=\linewidth]{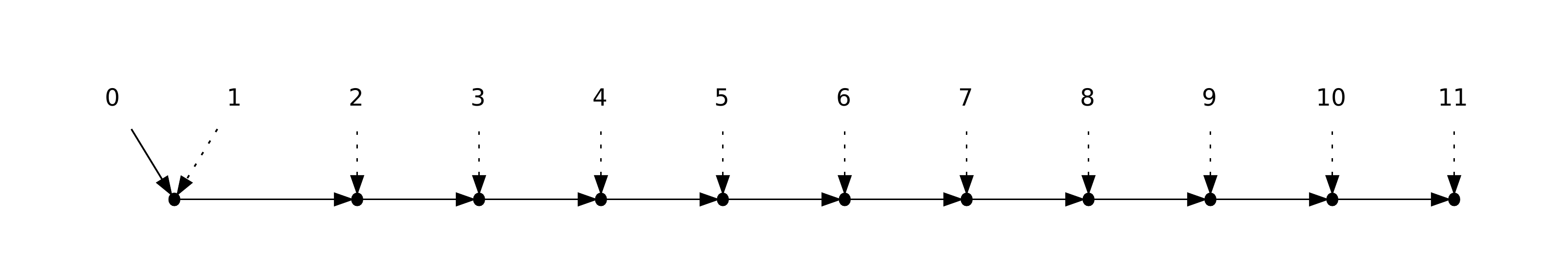}
\caption{Flat tree for panel $0$.}
\label{fig.tree.flat}
\end{figure}

Consider a panel with $m=12$.
All tiles except the diagonal, tile $0$, must be zeroed out.
We also know that in all algorithms, tile $0$ will be used as the killer in the last elimination.
The simplest solution is to use a single killer for the whole panel. If we do so, this single killer has to be the diagonal tile.
The eliminations will be all sequentialized (because the killer tile is modified  during each elimination),
but they can be performed in any order.
In Table~\ref{tab.panel.flat}, we use an ordering from top to bottom.
For each tile, we give the index of its killer. We also give the step at which it is zeroed out, assuming that each elimination can
be executed within one time unit.
The elimination list is then $\elim(0,1,0), \elim(0,2,0), \dots, \elim(0,m-1,0)$.
The corresponding \emph{reduction tree} for panel $0$ is a tree
with $m$ leaves and
$m-1$ internal nodes, one per elimination. Each internal node can also be viewed as the value of the killer tile just after
the elimination. Each internal node has two predecessors, namely the two tiles used to perform
the elimination.  In the example, internal nodes are arranged along a chain, with original tiles being input sequentially,
see Figure~\ref{fig.tree.flat}. The tree of Figure~\ref{fig.tree.flat} is called the \emph{flat tree}.
A tile eliminated at step $i$ in Table~\ref{tab.panel.flat} is at distance $S-i+1$ of the tree root,
where $S$ is the last time-step ($S=11$ in the example). Note that the reduction tree fully characterizes the elimination list
for the panel, since it provides both the killer and the time-step for each elimination.

\begin{table}[htb]
\centering
\begin{tabular}{|r|r|r|}%
\hline%
\multicolumn{1}{|c|}{Row index} & \multicolumn{1}{|c|}{Killer} & \multicolumn{1}{|c|}{Step} \\ \hline%
0 & \s &    \\
1 & 0 &  1  \\
2 & 0 &  2  \\
3 & 0 &  3  \\
4 & 0 &  4  \\
5 & 0 &  5  \\
6 & 0 &  6  \\
7 & 0 &  7  \\
8 & 0 &  8  \\
9 & 0 &  9  \\
10 & 0 &  10  \\
11 & 0 &  11  \\
\hline%
\end{tabular}%
  \caption{Flat tree reduction of panel $0$.}
  \label{tab.panel.flat}
\end{table}

\begin{figure}[htb]
\includegraphics[width=\linewidth]{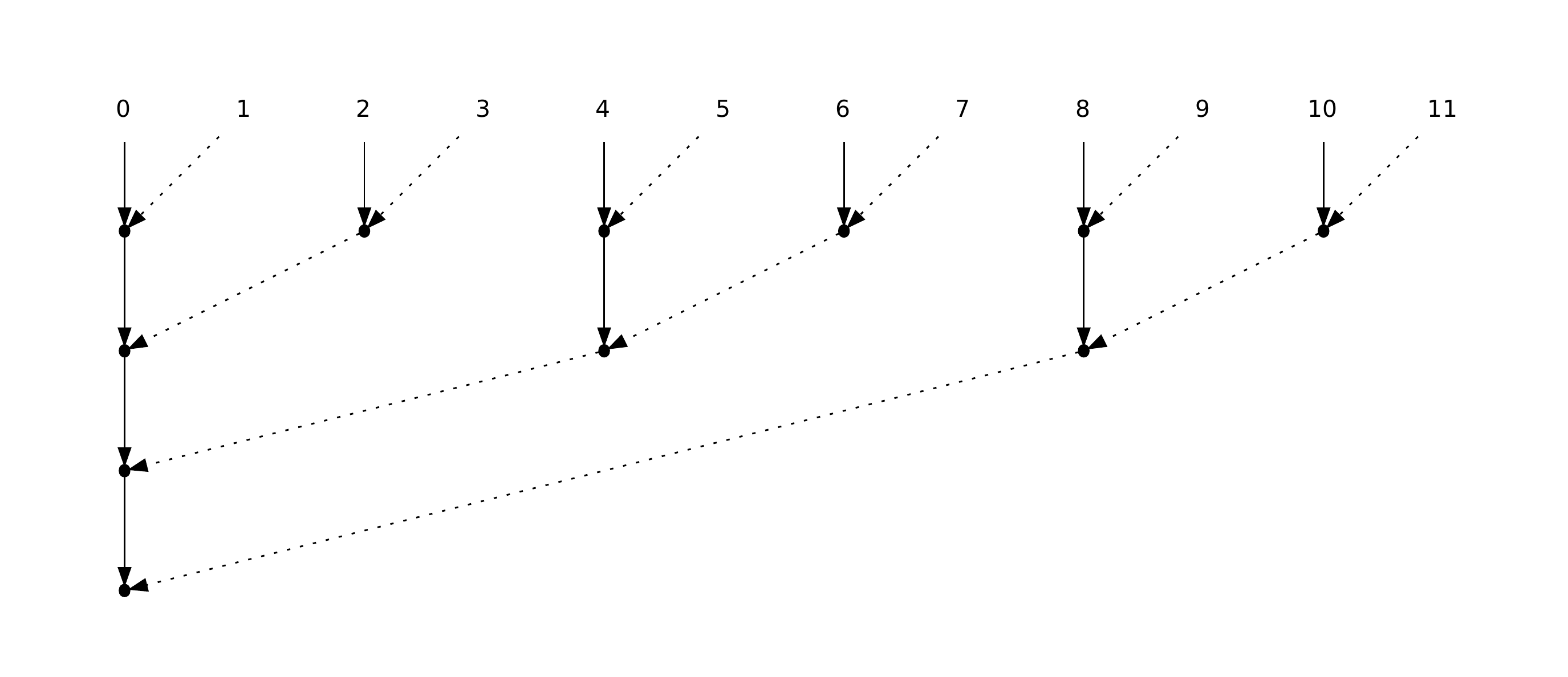}
\caption{Binary tree for panel $0$.}
\label{fig.tree.bin}
\end{figure}

With a single killer, all eliminations in the panel must be executed one after the other. The only source of parallelism resides
in the possibility to execute the updates of some previous eliminations while zeroing out the next tile. However,
parallel eliminations are possible
if we conduct these on disjoint pairs of rows. In the beginning, we can have as many killers as half the number of rows.
And the next step, half of the remaining non-zero rows can be killed. Iterating, we reduce the panel
with a \emph{binary tree} instead of a flat tree, as illustrated in  
Figure~\ref{fig.tree.bin}.
The elimination list is $\elim(2i,2i+1,0), i=0,\dots,\lceil \frac{m-2}{2} \rceil$, 
followed by $\elim(4i,4i+4,0), i=0,\dots,\lceil \frac{m-5}{4}\rceil$, and so on. The
last elimination is $\elim(0,2^{\lceil \log_2 m \rceil - 1},0)$.


With several killers, we have to use \TT elimination kernels, which are less efficient than \TS kernels.
This relative inefficiency of the \TT kernels is a first price to pay for parallelism.
A second price to pay arises from locality issues. In a shared-memory environment, re-using the same killer several times
allows for better cache reuse. This is even more true in distributed-memory environments, where the cost of communications can be much higher
than local memory accesses. In such environments, we have to account for the data distribution layout. 
Assume that we have
$p=3$ clusters $P_0$, $P_1$ and $P_2$. Here we use the term \emph{cluster} to denote either a single processor, or 
a shared-memory machine equipped with several cores. There are two classical ways to distribute rows to clusters, by blocks, or cyclically.
(In the general case one would use a 2D grid, but we use a 1D grid for
simplicity in this example).
These two distributions are outlined below:

\begin{center}
\scalebox{0.9}{
\begin{tabular}{|c||l|l|}
  \hline
  Clusters & Matrix rows (block) & Matrix rows (cyclic)\\ \hline
  $P_0$ & 0, 1, 2, 3 & 0, 3, 6, 9\\
  $P_1$ & 4, 5, 6, 7 & 1, 4, 7, 10\\
  $P_2$ & 8, 9, 10, 11 & 2, 5, 8, 11\\ \hline
\end{tabular}
}
\end{center}

In our example, the block distribution nicely fits with the flat tree reduction.
With this combination of block/flat, the ordering of the eliminations is such that the diagonal tile is communicated only once
from one cluster to the next one. Adding a last communication to store the tile back in $P_0$ gives a count
of $p$ communications. On the contrary, the cyclic distribution is communication-intensive for the flat tree reduction,
since we
obtain as many as $m$ communications, one per elimination and one for the final storage operation.
However, there are two important observations to make:
\begin{enumerate}
  \item With any data layout, one can always re-order the eliminations so as to perform
only $p$ communications with a flat tree. The killer can perform all local eliminations before
being sent to the next cluster. With the cyclic/flat combination in the example, we eliminate rows $3$, $6$, $9$,
then rows $1$, $4$, $7$, $10$, and finally rows $2$, $5$, $8$, $11$.
  \item There is a downside to fewer communications, namely higher start-up times. The cyclic/flat combination
  enables each cluster to become active much earlier (starting the updates) while the re-ordering dramatically increases waiting times.
  Note that waiting times are also high for the block/flat combination.
\end{enumerate}


\begin{figure}[!htb]
\includegraphics[width=\linewidth]{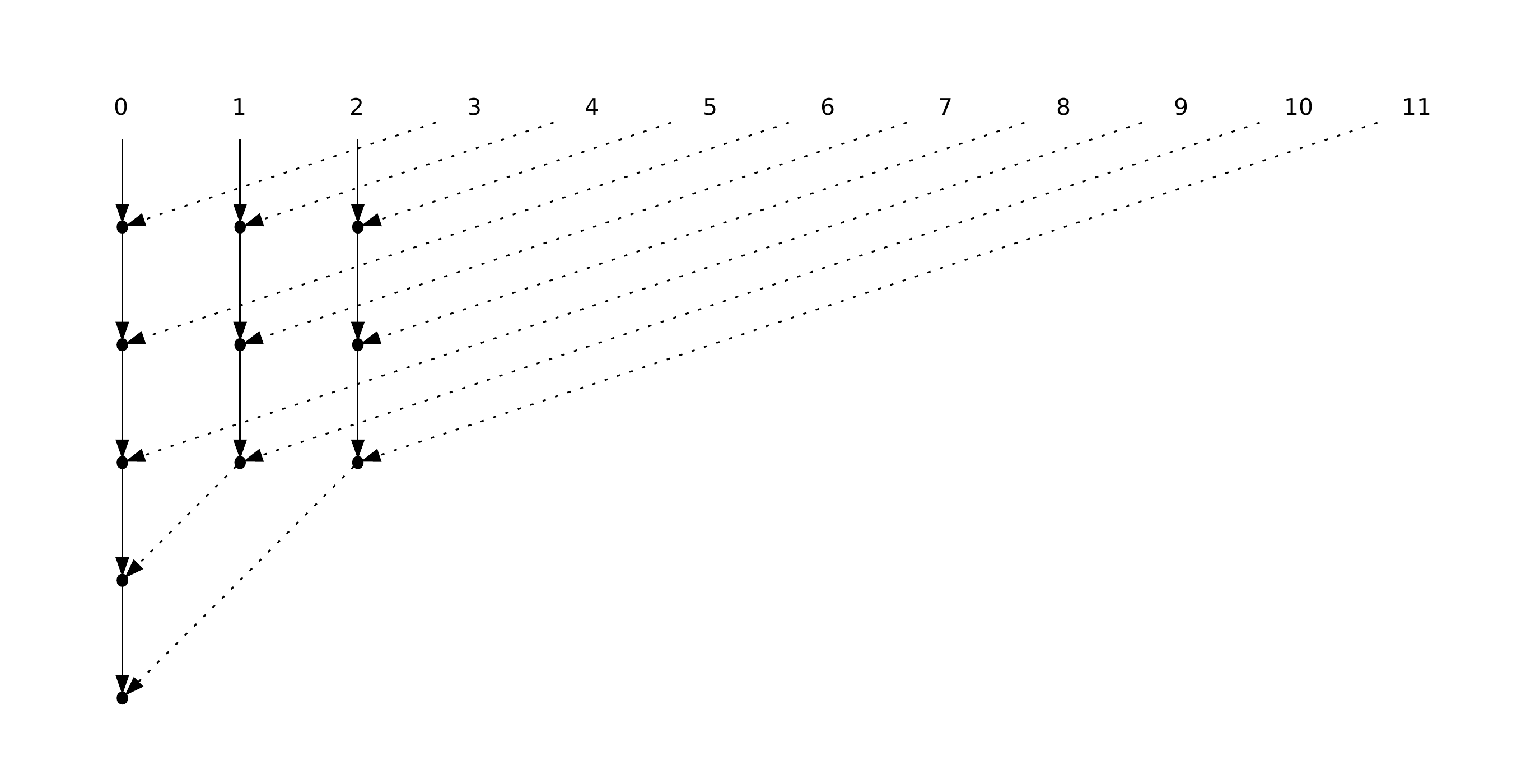}
  \caption{Flat/binary tree for panel $0$.}
  \label{fig.tree.flatbin}
\end{figure}

We make similar observations for the binary tree. In the example, the cyclic distribution requires many more
inter-cluster eliminations than the block one, which requires only two, namely the last two eliminations.
But this is an artifact of the example (take $p=4$ instead of $p=3$ to see this). In fact, for both distributions,
a better solution may be to use \emph{local} flat trees: within each cluster, a single tile acts as the killer for
all the local tiles. These flat trees are independent from one cluster to another, and the eliminations proceed
in parallel. Then a binary tree of size $p$ is used to eliminate $p-1$ out of the $p$ remaining tiles (one per cluster).
Communications are then reduced to a minimum. And because the local trees operate in parallel, there are no more high waiting
times at start-up, contrarily to the case with a single global killer given priority to local tiles.
This flat/binary reduction is illustrated in 
Figure~\ref{fig.tree.flatbin}.
In this example, the local killers are rows $0$, $1$ and $2$, and the binary tree has only $3$ leaves, one per cluster. 
Note that the tree is designed
with a cyclic distribution in mind: with a block distribution, the local killers would be rows $0$, $4$ and $8$.

\begin{figure}[!htb]
\includegraphics[width=\linewidth]{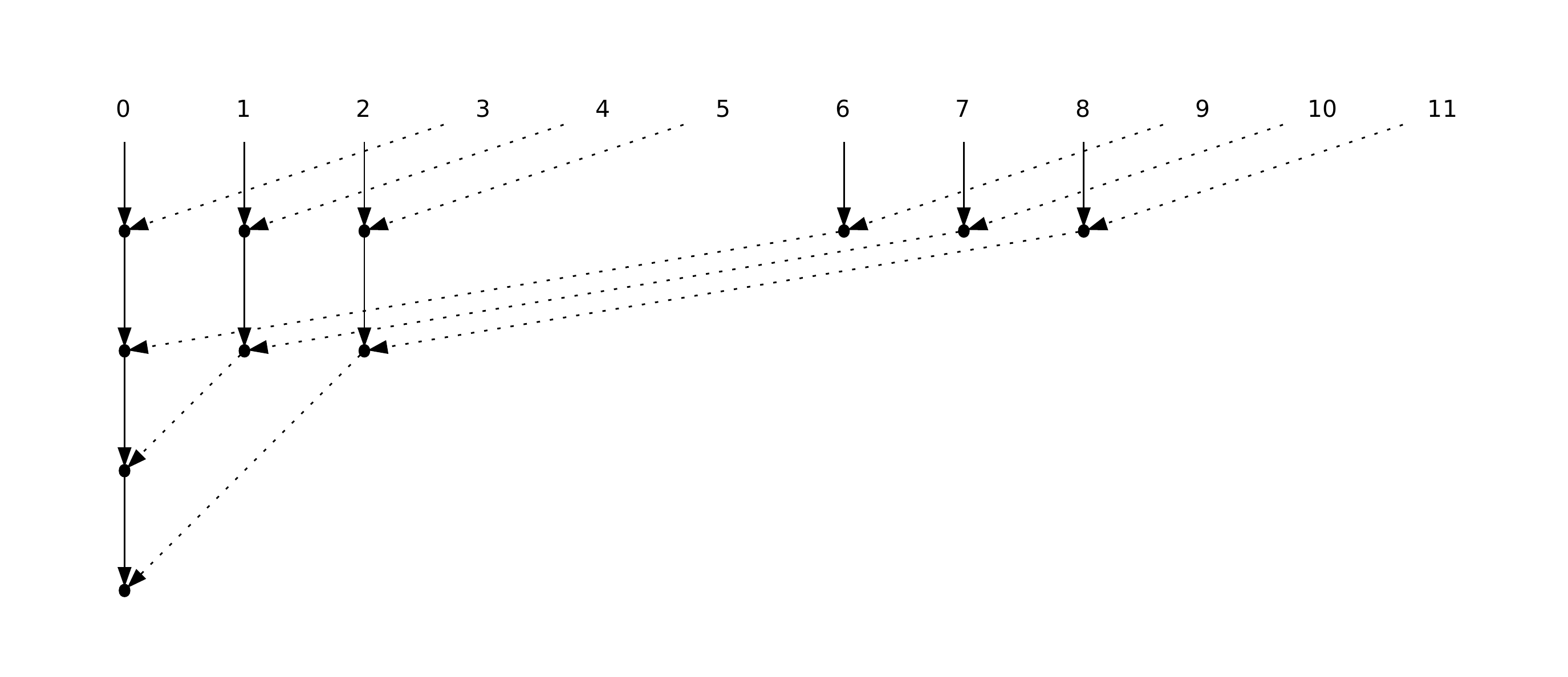}
  \caption{Domain tree for panel $0$, with two domains per cluster.}
  \label{fig.tree.domain}
\end{figure}

Further refinements can be proposed. The flat/binary strategy may suffer from not exhibiting enough parallelism at the
cluster level: local trees do execute in parallel, but each with a single killer. Parallelizing local eliminations
may be needed when the cluster is equipped with many cores.
The idea is then to partition the rows assigned to each cluster
into smaller-size \emph{domains}. Each domain is reduced using a flat tree, but there are more domains than clusters.
This domain tree reduction is illustrated in Figure~\ref{fig.tree.domain} with two domains
per cluster. In the example each domain is of size $2$, hence the corresponding flat tree is reduced to a single elimination,
but there are two domains, hence two killers, per cluster. The next question is: how to reduce these six killers?
We can  use a binary tree, as shown in Figure~\ref{fig.tree.domain}. But there is a lot of flexibility here. For instance
we may want  to give priority to local eliminations, hence to reduce locally 
before going inter-cluster. This amounts to using a local reduction tree to eliminate all domain killers but one within a cluster, and then a global reduction tree to eliminate all remaining killers but one within the panel. Let $m =p \times d \times a$, where 
$a$ is the domain size and $p$ the number of clusters. There are $d$ domains per cluster, hence each local reduction tree is of size $d$,
while the global reduction tree is of size $p$. Note that these two trees may well be of different nature, all combinations are allowed! 
In the example, there are only $d=2$ domains per cluster, so the local tree is unique, and using a binary tree for the global tree
leads to the same elimination scheme as using a single  binary tree for the six killers, as in Figure~\ref{fig.tree.domain}.

\subsection{Factoring several panels}

We have reviewed several strategies to factor the first panel of the $m \times n$ tile matrix. But the whole game amounts to factoring
$\min(m,n)$ panels, and efficiently pipelining these factorizations is critical to the performance of the QR algorithm.
This section aims at illustrating several trade-offs that can be made.

A striking observation is that using a flat tree reduction in each panel provides a perfect pipelining, while using a binary tree reduction
in each panel provokes ``bumps'' in the schedule, as illustrated with $3$ panels in Tables~\ref{tab.3panel.flat} and~\ref{tab.3panel.bin}.
This explains that flat trees have been predominantly used in the literature, until the advent of machines equipped with several cores.
Such architectures
called for using several killers in a given panel, hence for binary trees, and later domain trees.

\begin{table}[htb]
\centering
\begin{tabular}{|r||r|r||r|r||r|r|}%
\hline%
\multicolumn{1}{|c||}{Row index} & \multicolumn{2}{|c||}{Panel 0} & \multicolumn{2}{|c||}{Panel 1} & \multicolumn{2}{|c|}{Panel 2} \\
\multicolumn{1}{|c||}{ } & \multicolumn{1}{|c|}{Killer} & \multicolumn{1}{|c||}{Step} & \multicolumn{1}{|c|}{Killer} & \multicolumn{1}{|c||}{Step} & \multicolumn{1}{|c|}{Killer} & \multicolumn{1}{|c|}{Step}\\ \hline%
0 & \s &  & \s & & \s &  \\
1 & 0 &  1 & \s &  & \s & \\
2 & 0 &  2 & 1 & 3 & \s & \\
3 & 0 &  3 & 1 & 4 & 2 & 5\\
4 & 0 &  4 & 1 & 5 & 2 & 6\\
5 & 0 &  5 & 1 & 6 & 2 & 7\\
6 & 0 &  6  & 1 & 7 & 2 & 8\\
7 & 0 &  7  & 1 & 8 & 2 & 9\\
8 & 0 &  8  & 1 & 9 & 2 & 10\\
9 & 0 &  9  & 1 & 10 & 2 & 11\\
10 & 0 &  10  & 1 & 11 & 2 & 12\\
11 & 0 &  11  & 1 & 12 & 2 & 13\\
\hline%
\end{tabular}%
  \caption{Flat tree reduction for the first $3$ panels.}
  \label{tab.3panel.flat}
\end{table}

\begin{table}[htb]
\centering
\begin{tabular}{|r||r|r||r|r||r|r|}%
\hline%
\multicolumn{1}{|c||}{Row index} & \multicolumn{2}{|c||}{Panel 0} & \multicolumn{2}{|c||}{Panel 1} & \multicolumn{2}{|c|}{Panel 2} \\
\multicolumn{1}{|c||}{ } & \multicolumn{1}{|c|}{Killer} & \multicolumn{1}{|c||}{Step} & \multicolumn{1}{|c|}{Killer} & \multicolumn{1}{|c||}{Step} & \multicolumn{1}{|c|}{Killer} & \multicolumn{1}{|c|}{Step}\\ \hline%
 0 & \s  &    & \s &    & \s & \\
 1 &  0  &  1 & \s &    & \s & \\
 2 &  0  &  2 &  1 &  3  & \s & \\
 3 &  2  &  1 &  1 &  4  &  2  & 5\\
 4 &  0  &  3 &  3 &  4  &  2  & 7\\
 5 &  4  &  1 &  1 &  5  &  4  & 6\\
 6 &  4  &  2 &  5 &  3  &  2  & 9\\
 7 &  6  &  1 &  5 &  4  &  6  & 5\\
 8 &  0  &  4 &  7 &  5  &  6  & 8\\
 9 &  8  &  1 &  1 &  6  &  8  & 7\\
10 &  8  &  2 &  9 &  3  &  2  & 10\\
11 & 10  &  1 &  9 &  4  &  10  & 5\\
\hline%
\end{tabular}%
  \caption{Binary tree reduction for the first $3$ panels.}
  \label{tab.3panel.bin}
\end{table}

The inefficient pipelining of binary trees has only been identified recently. To remedy this problem while keeping several killers inside a panel,
one can use the \Greedy reduction outlined in Table~\ref{tab.3panel.greedy}.
The \Greedy algorithm nicely combines intra-panel parallelism and
inter-panel pipelining. In fact, under the simplifying assumption of unit-time eliminations (hence regardless of their number of updates),
it has been shown~\cite{j12,j14} that no algorithm can proceed faster!
At each step, the \Greedy algorithm eliminates as many tiles as
possible in each column, starting with bottom rows. The pairing for the
eliminations is done as follows: to kill a bunch of $z$ consecutive tiles
at the same time-step, the algorithm uses the $z$ rows above them as killers, pairing them
in the natural order. For instance in Table~\ref{tab.3panel.greedy}, the bottom six tiles in column $1$ are  simultaneously
killed during the first step, using the six tiles above them as killers.

\begin{table}[htb]
\centering
\begin{tabular}{|r||r|r||r|r||r|r|}%
\hline%
\multicolumn{1}{|c||}{Row index} & \multicolumn{2}{|c||}{Panel 0} & \multicolumn{2}{|c||}{Panel 1} & \multicolumn{2}{|c|}{Panel 2} \\
\multicolumn{1}{|c||}{ } & \multicolumn{1}{|c|}{Killer} & \multicolumn{1}{|c||}{Step} & \multicolumn{1}{|c|}{Killer} & \multicolumn{1}{|c||}{Step} & \multicolumn{1}{|c|}{Killer} & \multicolumn{1}{|c|}{Step}\\ \hline%
 0 & \s  &    & \s &    & \s & \\
 1 &  0  &  4 & \s &    & \s & \\
 2 &  1  &  3 &  1 &  6  & \s & \\
 3 &  0  &  2 &  2 &  5  &  2  & 8\\
 4 &  1  &  2 &  2 &  4  & 3  & 7\\
 5 &  2  &  2 &  3 &  4  & 4   & 6\\
 6 &  0  &  1 &  3 &  3  & 5   & 6\\
 7 &  1  &  1 &  4 &  3  & 5   & 5\\
 8 &  2  &  1 &  5 &  3  & 6  & 5\\
 9 &  3  &  1 &  6 &  2  & 7   & 4\\
10 &  4  &  1 &  7 &  2  & 8  & 4\\
11 & 5  &   1 &  8 &  2  & 10  & 3\\
\hline%
\end{tabular}
  \caption{Greedy reduction for the first $3$ panels.}
  \label{tab.3panel.greedy}
\end{table}

Recall from the study with a single panel that locality issues are very important in a distributed-memory environment, i.e. with several clusters. The previous \Greedy algorithm is not suited to a matrix whose rows have been distributed across clusters, and two levels
of reduction, local then global, are still highly desirable. But in addition to locality, a new issue arises
when factoring a full matrix instead of a single panel: because the number of active rows decreases from one panel to the next, block distributions
are no longer equivalent to cyclic distributions: the former induces a severe load imbalance (clusters become inactive as the execution progresses)
while the latter guarantees that each cluster receives a fair share of the work until the very end of the factorization.

Finally, we point out that dealing with a coarse-grain model where each elimination requires one time unit, 
as in all previous tables and figures, is a drastic simplification. Tiled algorithms work at the tile level: after each zero-ing out, 
as many update tasks are generated as there are trailing columns after the current panel. The total number of tasks that are created 
during the algorithm is
proportional to the cube of the number of tiles, and schedulers must typically set priorities to decide which tasks to execute among those
ready for execution. Still, the coarse-grain model allows us to understand the main principles that guide the design of tiled QR algorithms.

\subsection{Existing tiled QR algorithms}
\label{sec.related}

While the advent of multi-core machines is somewhat recent, there is a long line of papers related to tiled QR
factorization.
Tiled QR algorithms have first been
introduced in~Buttari et al.~\cite{Buttari2008,tileplasma} and Quintana-Ort{\'\i} et al.~\cite{Quintana:2009}
for shared-memory (multi-core) environments, with an initial focus on square matrices. The sequence of eliminations presented
in these papers is analogous to \SK~\cite{SamehKuck78}, and corresponds to reducing each panel with a flat tree:
in each column, there is a unique killer, namely the diagonal tile.

The introduction of several killers in a given column dates back to~\cite{SamehKuck78,ModiClarke84,Pothen89},
although in the context of traditional element-wise (non-blocked) algorithms. 

In the context of a single tile column, the first use of a binary tree algorithm (working on tiles) is due to
da Cunha et al.~\cite{Cunha2002}. Demmel et al.~\cite{CAQR} present a general tile algorithm where any tree can
be used for each panel, while Langou~\cite{me} explains the tile panel factorization as a reduction operation.

For shared-memory (multi-core) environments,  recent work advocates the use of domain 
trees~\cite{Hadri_ipdps_2010} to expose more parallelism with several killers while enforcing
some locality within domains.   Another recent paper~\cite{sc-paper2011} introduces tiled versions of the Greedy 
algorithm~\cite{j12,j14} and of the Fibonacci scheme~\cite{ModiClarke84}, and shows that these algorithms are asymptotically optimal. In addition, they experimentally turn out to outperform all previous algorithms for tall and skinny matrices.

Preliminary hierarchical two-level trees has been presented by Agullo et al. in the context of grid computing environment~\cite{tsqr-grid}
(binary on top of binary, for tall and skinny matrices), Agullo et al. in the context of multicore platform~\cite{Hadri_ipdps_2010}
(binary on top of flat, for any matrix shapes), and Demmel et al. in the context of multicore platform~\cite{mohiyuddin:SC09}
(binary on top of flat, for tall and skinny matrices).

In this paper,
we further investigate the impact of the Greedy and Fibonacci schemes, but for distributed-memory environments.
There are two recent works for such environments. The approach of~\cite{tsqr-grid} uses a hierarchical approach:
for each matrix panel, it combines two levels of reduction trees: first several local binary trees are applied in parallel,
one within each cluster, 
and then a global binary tree is applied for the final reduction across clusters. Because ~\cite{tsqr-grid} focuses on
tall and skinny matrices, it uses a 1D block distribution for the matrix layout (hence a 1D cluster grid).
The approach of~\cite{SongLHD10} also uses a hierarchical approach, and also uses a 1D block distribution. The main
difference is that the first level of reduction is conducted with a flat tree within each cluster. We point out
that the block distribution is suited only for tall and skinny matrices, not for general matrices. Indeed, with an
$m \times n$ matrix and $p$ clusters, the cyclic distribution is perfectly balanced (neglecting lower order terms),
while the speedup attainable by the block distribution is bounded by $p(1 - \frac{n}{3m})$: this is acceptable if
$n \ll m$ but a high price to pay if, say, $m=n$. However, it is quite possible to modify the algorithm in~\cite{SongLHD10}
so as to use a cyclic distribution, at the condition of re-ordering the eliminations to give priority to local ones
over those that require inter-cluster communications.  In fact, the hierarchical algorithm introduced in this paper
can be parametrized to implement either version, the original algorithm in~\cite{SongLHD10} as well as the latter variant
with cyclic layout.


\section{Hierarchical algorithm}
\label{sec.newalgo}

This section is devoted to the new hierarchical algorithm that we introduce for clusters of multicores. We outline 
the general principles (Section~\ref{sec.generaldesc-algo}) before working out the technical details through an 
example (Section~\ref{sec.example-algo}). Then we 
briefly discuss the  implementation within the \Dague framework in Section~\ref{sec.dague}.

\subsection{General description}
\label{sec.generaldesc-algo}

Here is a high-level description of the features of the  hierarchical
algorithm, \hqr: 
\begin{itemize}
   \item Use a 2D cyclic distribution of tiles along a virtual $p \times q$ cluster grid. The 2D-cyclic 
   distribution is the one that best balances the load across resources.
    \item Use domains of $a$ tiles,  and use  \TS kernels within domains. 
    Thus, within each cluster, every $a$-th tile sequentially kills the $a-1$ tiles  below it. The idea is to benefit
    from the arithmetic efficiency of \TS kernels.  Note that if $a=1$, the algorithm will use only \TT kernels.
  \item Use intra-cluster reduction trees within clusters. Here, the idea is to locally kill as many tiles as possible, without inter-processor communication. These intra-cluster trees depend upon the internal degree
  of parallelism of the clusters: we can use a binary tree or a \Greedy reduction for clusters with many cores, or a flat tree reduction
  if more locality and CPU efficiency is searched for. Note that these reductions are necessarily based upon \TT kernels,
  because they involve killer tiles from the domains.
  \item Use inter-cluster reduction trees across clusters (again, necessarily based upon \TT kernels).
  The inter-cluster reduction trees are of size $p$, because for each panel they involve a single tile per cluster.
  Here also, the trees can be freely chosen (flat, binary, greedy).
\end{itemize}

There are many parameters to explore: the arithmetic performance parameter $a$, the shape $p \times q$ of the 
virtual grid
if we are given $C_1$ physical clusters with $C_2$ cores each, and the shape of the intra- and inter- cluster reduction trees. 
In fact, there are two additional complications:
\begin{itemize}
  \item Consider a given cluster: ideally, we would like to kill all tiles but one in each panel, i.e., we would like to reduce each cluster sub-matrix to a diagonal, and then proceed with inter-cluster communications to finish up the elimination. Unfortunately, 
  because of the updates, it is not possible to locally kill ``in advance'' so many tiles, and one needs to wait for the inter-processor reduction to progress significantly to be able to perform the last local eliminations.  This scheme is explained in
  Section~\ref{sec.example-algo} below.
 
%
%

  \item The actual (physical) distribution of tiles to clusters needs not obey the virtual $p \times q$ cluster grid. In fact, we can
  always use another grid to map tiles to processors. This additional flexibility allows us to execute all previously
  published algorithms simply by tuning the actual distribution parameters. For instance, to run the algorithm
  of~\cite{SongLHD10} on a $m \times n$ tiled matrix, using a block distribution on $r$ processors, 
  we take a virtual grid value $p=1$ 
  with domains of size $a=m/r$, and  we let the actual data distribution be $\CYCLIC(r)$.
\end{itemize}


\begin{figure}[!htbp]
  \begin{center}
    \subfigure[Global]{
      \includegraphics[width=0.45\linewidth]{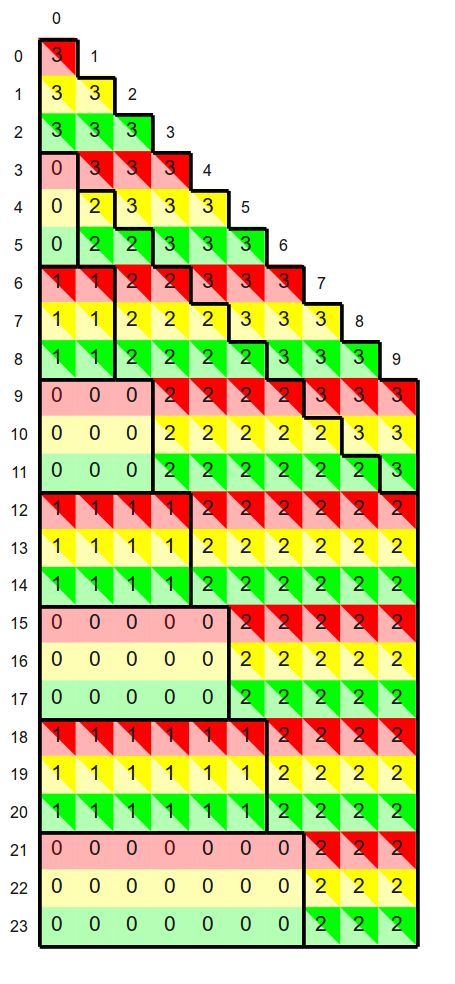}
      \label{fig:TypeMatGlobal}
    }
    \subfigure[Local]{
      \includegraphics[width=0.45\linewidth]{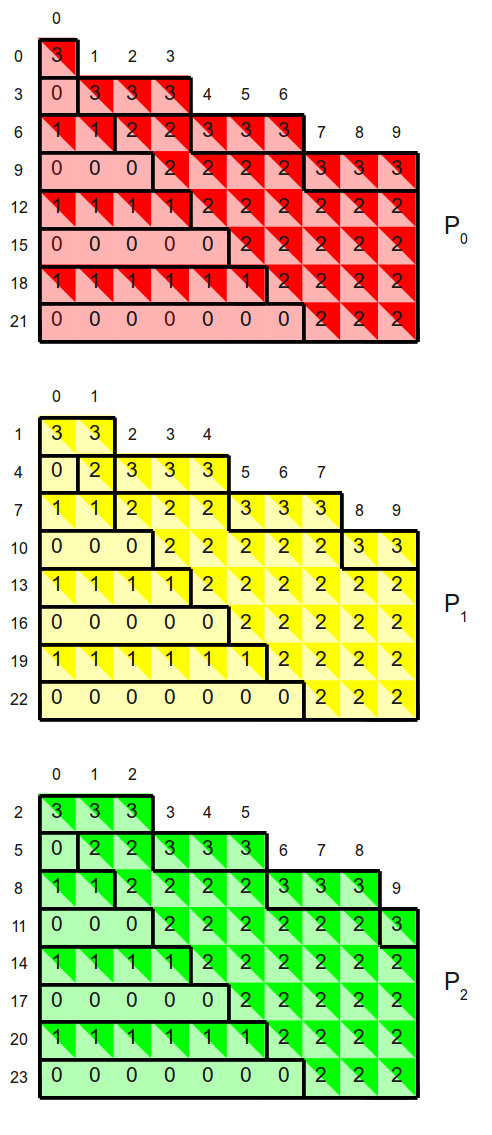}
      \label{fig:TypeMatLocal}
    }
  \end{center}
  \caption{Views of the tile labels.}
\end{figure}


\subsection{Working out an example}
\label{sec.example-algo}

Consider a $m \times n$ tiled matrix, with $m=24$ and $n=10$. We use a $p \times q$ virtual grid with $p=3$ and $q=1$,
and an arithmetic parameter $a=2$. Thus we have a unidimensional grid
with $p=3$ clusters. A global view of the matrix is given in Figure~\ref{fig:TypeMatGlobal},
while local distributions within each cluster are shown in Figures~\ref{fig:TypeMatLocal}. In both figures,
 tiles are colored according to their assigned processor (red for $P_{0}$,
yellow for $P_{1}$ and green for $P_{2}$). The label inside each tile characterizes its level of reduction, as explained
below.

\paragraph{Level $0$ tiles--} we have domains of size $a=2$, so that in essence every second tile is killed by a \TS kernel,
and the killer is always the tile above it in the local view of the figure~\ref{fig:TypeMatLocal}. However, as shown
in Figure~\ref{fig:TypeMatLocal}, this holds true
only for even-numbered tiles that are below the local diagonal. This local diagonal is a line of slope $1$ in the local view,
hence of slope $p$ in the global view.
If the matrix is tall and skinny, the proportion of level $0$ tiles
tends to be one half, but it is much less for square matrices.

\paragraph{Level $1$ tiles--} level $1$ tiles are the local killers of level $0$ tiles that lie strictly below 
the local diagonal. Such tiles can be killed locally, without any inter-cluster communication. In other words,
it is possible to kill all tiles of level $0$ and $1$ locally, in parallel on each cluster, before needing any inter-cluster communication.
At the end of this local elimination, all tiles lying in the lower triangle below the local diagonal have been killed,
and the last killer on each panel is the tile on the local diagonal (e.g., tile $(6,2)$ for panel $2$ in cluster $P_{0}$).
The elimination of the lower triangle can be conducted using various types of reduction trees, flat, binary or \Greedy.

\paragraph{Level $3$ tiles--} consider the panel of index $k$, and a cluster $P_{q}$. Consider the top tile on or  
below the matrix diagonal, i.e., the first tile in column $k$ whose row index is at least $k$. If this tile has row index $k$,
it is the diagonal tile; otherwise, if its row index is greater than $k$, it will be the last tile
killed in this panel. There are $p$ such top tiles, one per cluster,  and they are located on the first $p$ diagonals of the matrix.
Reducing the $p$ top tiles for a given panel induces inter-cluster communications. Within each panel, this
high-level reduction tree is of size $p$, and be freely chosen as flat, binary or \Greedy.

\paragraph{Level $2$ tiles--} these are the ``domino'' tiles. In each panel, using the local view within a cluster,
they are located between the top tile (not included) and the local diagonal tile (included). 
Their number increases together with the panel index, since level $2$ tiles lie between a line of slope $1/p$
and one of slope $1$ in the local view. While level $0$ and level $1$ tiles are killed independently within each cluster,
level $2$ tiles can only be killed after some inter-cluster communication has taken place. The goal of
the coupling level tree is to efficiently resolve interactions between local reductions and global reductions, and to kill
all level $2$ tiles as soon as possible.
To see the coupling level tree in action, consider the first level $2$ tile, in position $(4,1)$ and assigned to $P_{1}$. Tile $(4,1)$ is killed by
tile $(1,1)$, the top tile of $P_{1}$ for panel $1$: this corresponds to 
the elimination $\elim(4,1,1)$, which is intra-cluster (within $P_{1}$). 
But tile $(1,1)$ is not ready to kill tile $(4,1)$ until it has been
updated for the elimination $\elim(1,0,0)$, which is inter-cluster: level $3$ tile $(0,0)$ kills level $3$ tile $(1,0)$, 
and tile $(1,1)$ is updated during this elimination. As soon as the update ends, $\elim(4,1,1)$ is triggered, 
and tile $(4,1)$ is killed. A similar sequence takes place on to $P_{2}$, where the update of  tile $(2,1)$ during 
$\elim(2,0,0)$ (inter-cluster) must precede the killing of level 2 tile $(5,1)$ (during $\elim(5,2,1)$, intra-cluster).
In fact, we see that inter-cluster eliminations in the high-level tree successively trigger eliminations in the coupling tree,
like a domino that ripples in the area of level $2$ tiles. 

\paragraph{Execution scheme--} with an infinite number of resources, the execution would progress as fast as possible.
The elimination list of the algorithm is the composition of the reduction trees at all the different levels. All killers are known
before the execution. Each component of an elimination is triggered as soon as possible, i.e. as soon as all dependencies are satisfied: 
first we have the killing of the tile, and then the updates in the
trailing panels. Note that the overall elimination scheme is complex,
and mixes the killing of tiles at all levels. With a fixed number of
resources, it is necessary to decide an order of execution of the
tasks, hence to schedule them: this is achieved through the  \Dague environment.

\subsection{Implementation with \Dague}
\label{sec.dague}

\Dague is a high-performance fully-distributed scheduling environment
for systems of micro-tasks. It takes as input a
problem-size-independent, symbolic representation of a Direct Acyclic Graph of 
tasks, and schedules them at runtime on a distributed parallel machine of multi-cores. Data
movements are expressed implicitly by the data flow between the tasks
in the DAG representation. The runtime engine is then responsible for
actually moving the data from one machine (cluster) to another, using an underlying
communication mechanism, like MPI. A full description of \Dague, and
the implementation of classical linear algebra
factorizations in this environment, can be found in~\cite{dague-engine, dague-la}.

To implement the generic QR algorithm in \Dague, it is sufficient to
give an abstract representation of all the tasks (eliminations and
updates) that constitute the QR factorization, and how data flows from
one task to the other. Since a tiled QR algorithm is fully determined
by its elimination list, this basically consists only into providing a function
that the runtime engine is capable of evaluating, and that computes
this elimination list. The \Dague object obtained this way is generic:
when instantiating a \Dague QR factorization, the user sets all
parameters that define this elimination list ($p, q, a$, the shape of
the local and high-level trees), defining a new DAG at each
instantiation. 

At runtime, tasks executions trigger data movements, and create new
ready tasks, following the dependencies defined by the
elimination list. Tasks that are ready to compute are scheduled
according to a data-reuse heuristic: each core will try to execute close
successors of the last task it ran, under the assumption that these
tasks require data that was just touched by the terminated one. This
policy is tuned by the user through a priority function: among the
tasks of a given core, the choice is done following this function. To
balance load between the cores, tasks of a same cluster in the
algorithm (reside on a same shared memory machine) are shared between
the computing cores, and a NUMA-aware job stealing policy is
implemented. The user is responsible for defining the affinity between
data and tasks, and to distribute the data between the computing
nodes. Thus, she defines which task execute on which node, and
remains responsible for this level of load balancing. In our case, the
data distribution is a $p \times q$ grid of $b \times b$ tiles, with
a cyclic distribution \CYCLIC( 1 ) of tiles across both grid dimensions.

\section{Experiments}
\label{sec.expe}

\begin{figure*}
\begin{center}
\subfigure[low-level tree set to Greedy\label{fig:a-histo-greedy}]{\includegraphics[width=.45\linewidth]{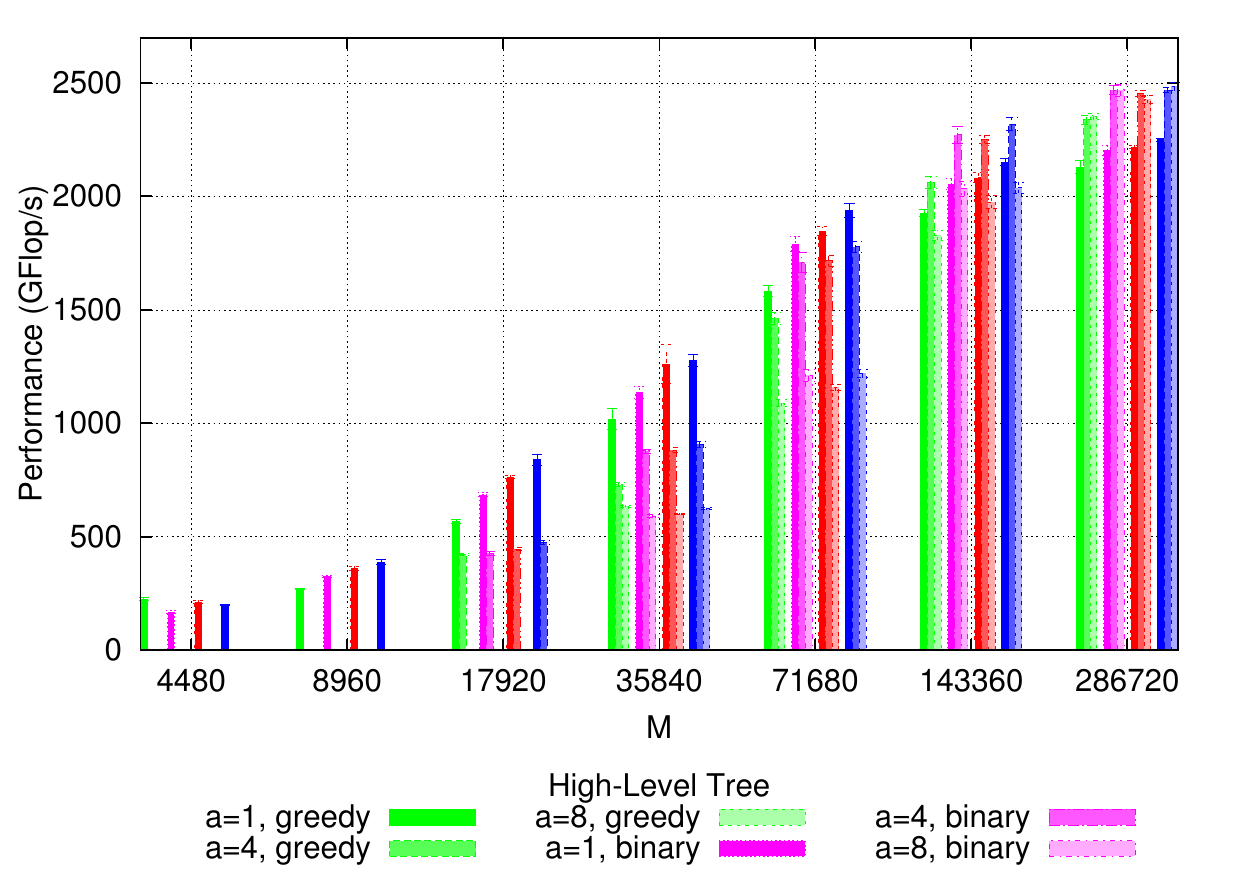}}
\subfigure[low-level tree set to Flat\label{fig:a-histo-flat}]{\includegraphics[width=.45\linewidth]{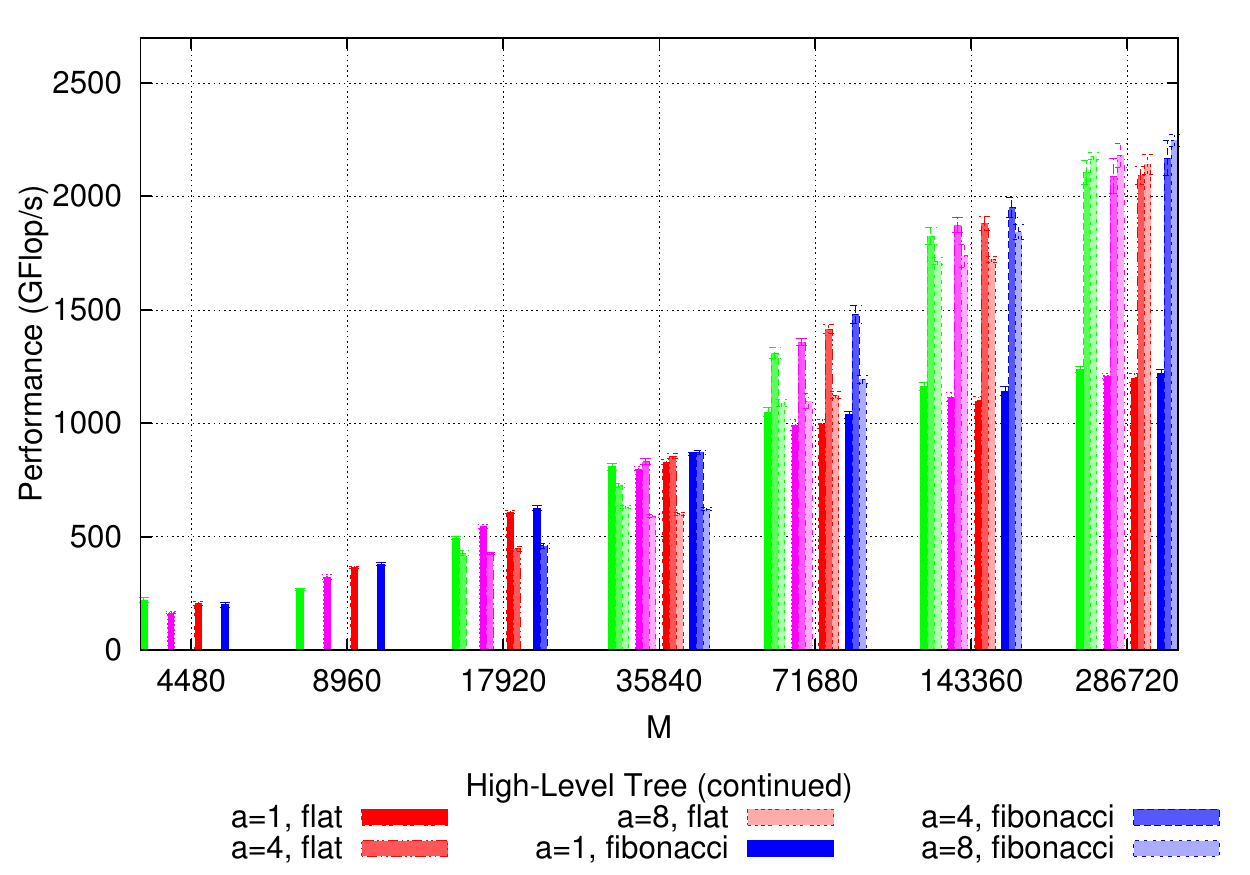}}
\caption{Performance of the \hqr algorithm on a $M\times 4,480$
  matrix. (Domino optimization not activated). Influence of the TS level ($a$ value), low level and high level trees.
  \label{fig:a-histo}}
\end{center}
\end{figure*}

\subsection{Experimental Conditions}

The purpose of this performance evaluation is to highlight the
features of the proposed algorithm, and to compare its efficiency with
state-of-the-art QR factorization implementations.
We use \emph{edel}, a parallel machine hosted by the Grid'5000 experimental
platform~\cite{grid5000}, to support the experiments. 
These experiments feature $60$ multi-core machines, each equipped with $8$ cores, and an Infiniband 20G
interconnection network. The machines feature two NUMA Nehalem Xeon E5520 at 2.27GHz
(hyperthreading is disabled), with 12GB of memory (24GB per machine). The
system is running the Linux 64bit operating system, version
2.6.32-5-amd64 (Debian 2.6.32-35). The software is compiled with Gcc
version 4.4.5, and GFortran 4.4.5 when applicable. BLAS kernels were
provided by the MKL library from the Intel compiler suite 11.1. The
\Dague software from the mercurial repository revision 3130 uses
Open~MPI version 1.4.3 as network backend. All experiments have been run
at least 5 times, and the average value is presented, together with the
standard deviation. We use {\em whiskers} to represent standard deviation on all of our figures.
For each experiment, we compute the Q factor of the QR factorization (by
applying the reverse trees to the identity) and check (a) that $Q$ has
orthonormal columns and (b) that $A$ is equal to $Q*R$.  All checks were satisfactory up
to machine precision.

The theoretical peak performance of this machine for double-precision
is 9.08 GFlop/s per core, 72.64 GFlop/s per node, and 4.358 TFlop/s
for the whole machine. The best performance for running the dTSMQR
operation in a single core, has been measured at 7.21 GFlop/s (79.4\%
of the theoretical peak), and the dTTMQR operation has been measured at
6.28 GFlop/s (69.2\% of the theoretical peak).
Depending on the $a$ value chosen, 
these numbers can be seen as practical peaks.
For example, if $a=1$, most of the flops are in dTTMQR (69.2\% of the theoretical peak). 
As $a$ gets larger, more flops shift to dTSMQR (79.4\%
of the theoretical peak).

Our implementation of \hqr operates on a virtual grid $p \times q$ set
to $15 \times 4$, it feature a TS level with parameter $a$ (set
$a$ to 1 for no TS, and $a=m/p$ for full \TS on the node), a choice of four different TT trees
for the low level (\Greedy, \BT, \FT, \MC), the coupling level can be activated
or not. When it is activated, the domino TT tree is used by default,
and there is a choice of
four different TT trees for the high level (\Greedy, \BT, \FT,
\MC). Tiles of size $b \times b$ are used. 
The \Dague engine offers several data distribution and automatically handles the data transfers when needed. 
As a consequence, our \Dague implementation would operate on any \Dague-supported data distribution.
For \hqr, we focus on 2D block cyclic distribution using a $p \times q$ process
grid mapping the algorithm virtual grid.

We compare our algorithm to [BBD+10]~\cite{dague-la},
[SLHD10]~\cite{SongLHD10}, and \scalapack~\cite{Scalapack97}.
Since [SLHD10] is a sub-case of the \hqr algorithm
(see Section~\ref{sec.generaldesc-algo}),
we use our \Dague-based implementation of \hqr to execute it.  [SLHD10] for a $m \times n$ tiled matrix, 
using a block distribution on $p$ processors, corresponds to the \hqr algorithm with the following parameters:
virtual grid value $p=1$, domains of size $a=m/p$, data distribution $\CYCLIC(a)$, low-level binary tree.
(Since $p=1$, neither the coupling level nor the high level are relevant.)
[BBD+10] corresponds to the QR operation currently available in \Dague,
which implements the Tile QR factorization described in~\cite{dague-la}.
%
%
%
\scalapack experiments use the \scalapack implementation of the QR
factorization found in the MKL libraries. The MKL number of threads was
set to $8$, and one MPI process was launched per node.
For all other setups (that are \Dague
based), the binary was linked with the sequential version of the MKL
library, and \Dague was launched with $8$ computing threads and an
additional communication thread per node. All threads are bound to a
different core, except the communication thread that is allowed to run
on any core.

In all experiments,
we used 60 nodes (480 cores), and the data was distributed
along a $15\times 4$ process grid for \hqr, [BDD+10], and \scalapack,
and a $60\times 1$ 1D block distribution
for [SLHD10]. 

All \hqr runs use a \emph{virtual} cluster grid exactly mapping the
\emph{process} grid used for data distribution. The coupling-tree, whenever
activated, is implemented with the so-called domino scheme.  We fix the tile
size parameter $b$ in our experiments as being the block size which renders the
best sequential performance for the sequential TS update kernel.  More tuning
could be done for \hqr with respect to the tile size and to the process grid
shape parameters. In particular, $b$ directly influences at least two key
performance metrics, namely the number of messages sent and the granularity of
the algorithm.  We have fixed these parameters for the whole experiment set.
Choosing $b=280$ and a process grid $p\times q$ of $15\times 4$ leads to values
that consistently provide good performance.




\subsection{Evaluation of \hqr}

\hqr is a highly modular algorithms. The design space offers by its parameters
is large. The goal of this section is to confront our intuition of \hqr with
experimental data in order to build up understanding on how these
parameters influence the overall performance of \hqr.  In
Section~\ref{sec:on.est.les.meilleurs}, we use this newly acquired understanding
to set up the parameters for various fixed-parameters experiments.  We note
that, overall, \hqr is an intrinsically better algorithm than what has been
proposed in the past.  Although we explain in this section that some significant
performance gains can be obtained by tuning the parameters, setting some
default values is enough to outperform the current state of the art.

\begin{figure}
\begin{center}
\includegraphics[width=.9\linewidth]{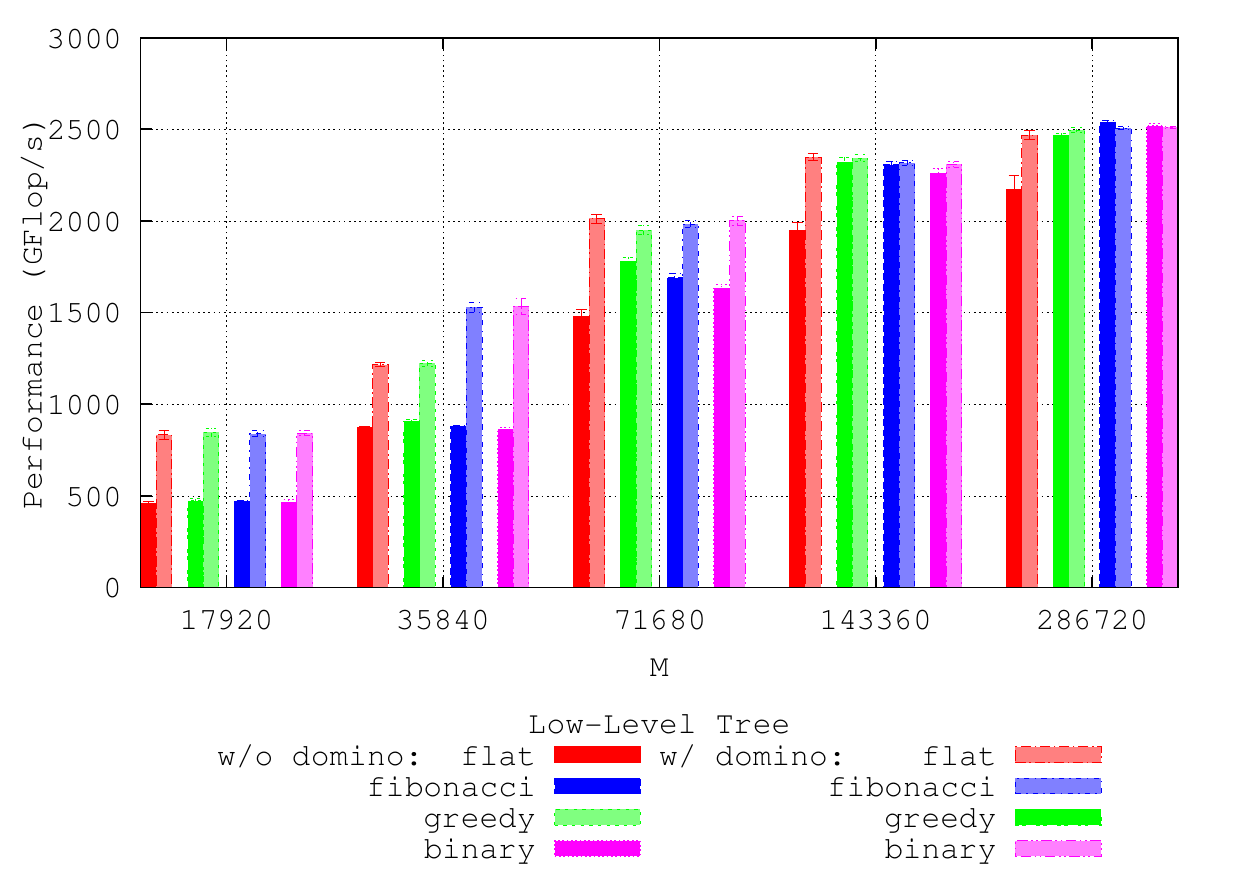}
\caption{Performance of the \hqr algorithm, on a $M\times 4,480$
  matrix. High-level set to
  Fibonacci and $a=4$. Influence of the low-level tree and the domino optimization. \label{fig:domino-histo}}
\end{center}
\end{figure}

Figure~\ref{fig:a-histo} presents the performance of the Hierarchical
QR algorithm, \hqr, for different matrix sizes, different trees and
different values of the $a$ parameter. The matrix size varies from a
square matrix of $16\times 16$ tiles to a tall and skinny matrix of
$1,024\times 16$ tiles. Since we are working on a $15\times 4$ process
grid, this means that local matrices range from $1\times 4$ tiles
to $68\times 4$ tiles.  
In order to first focus only on the influence of the TS level, low level and high level trees, the domino coupling optimization is not yet activated.
Subfigure~\ref{fig:a-histo-greedy} presents the performance for
all possible high-level trees with a low-level tree set to
\Greedy, while Subfigure~\ref{fig:a-histo-flat} presents
the same with a low-level tree set to \FT. Figures with a
low-level tree set to \BT or \MC are
omitted due to lack of space; however they exhibit a behavior similar
to Figure~\ref{fig:a-histo-greedy} (\Greedy).
Figure~\ref{fig:domino-histo} presents the performance of the
\hqr algorithm, for the same set of matrices, with a fixed
value $a=4$, and a high-level tree set to
\MC. Measurements were done alternatively turning on or off the
domino optimization presented in
Section~\ref{sec.example-algo}.

{\em Influence of $a$.} Looking at Subfigure~\ref{fig:a-histo-greedy}, we see
that, for small values of $M$, the value $a=1$ is best. This is because a
higher value of $a$ negatively impacts the degree of parallelism of the
algorithm when we use the low level \Greedy tree on small matrices.
When $M$ increases, the number of tasks increases, and we end up with
abundant parallelism.  Consequently, we can safely increase
the value of $a$ up to $4$ or $8$. For large $M$, we see that the speedup between
$a=1$ and $a=4,8$ is about 10\% which is the speedup between TT update kernels
and TS update kernels. When the low level tree is \FT,
(Subfigure~\ref{fig:a-histo-flat}), we have a different story. Adding a flat
tree (TS kernels) beneath a low-level flat tree in the tall and skinny case
(large $M$) actually increases the parallelism. In effect, the TS flat trees
divide the length of the pipeline created by the low level flat tree by a factor $a$.
So there are two benefits for tall and skinny matrices in adding a flat tree TS
beneath a flat tree TT: (1) faster kernels; (2) better parallelism.  This
explains why the speedup for $a=4$ or $a=8$ with respect to $a=1$ is way above
10\% for large $M$.  Altogether, we conclude that significant gain can be obtain by tuning the parameter $a$ for
various matrix shapes, number of processors and TT vs TS ratio. 

{\em Influence of the low level tree.} For tall and skinny matrices, \Greedy
is better than \FT. In the $286,720\times 4,480$ case, the
low level tree performs on a $68 \times 16$ matrix ($m/p \times n$), and in that case
the critical path length of flat tree is approximately 2.6x  the one of greedy (($(68+2*16
)/(\log_2(68)+2*16)$)~\cite{sc-paper2011}).  
Looking at Subfigures~\ref{fig:a-histo-greedy} and~\ref{fig:a-histo-flat}, we see a speedup
of about 2x when the low level tree changes from flat tree to greedy in the
$a=1$ case. When $a$ increases, the low level trees affect fewer tiles
and, consequently, its influence on the overall algorithm is reduced.
See also Figure~\ref{fig:domino-histo}, where we have set $a=4$, and we
observe that all low level trees perform more or less similarly.

{\em Influence of the high level tree.} We observe similar performances for all variants,
although Fibonacci is slightly better than its competitors.

{\em Influence of the coupling level tree (domino optimization).} In Figure~\ref{fig:domino-histo}, we see the
positive effect of the domino optimization in the case of tall and skinny matrices. When activated,
for a tall and skinny matrices, it
never significantly deteriorates the performance and can have significant
impact.
The domino optimization is all the more
important when a good coupling between the local tree and the distributed tree
is critical. This is illustrated best with the case of low level
\FT. Indeed, this optimization
enables look-ahead on the local panels as explained in
Section~\ref{sec.example-algo}, thereby increasing the degree of
parallelism.
Although not reported in this manuscript, we note that domino optimization have a negative impact when the matrix becomes large and square.

\subsection{Comparison}
\label{sec:on.est.les.meilleurs}

\begin{figure}
\begin{center}
\includegraphics[width=.9\linewidth]{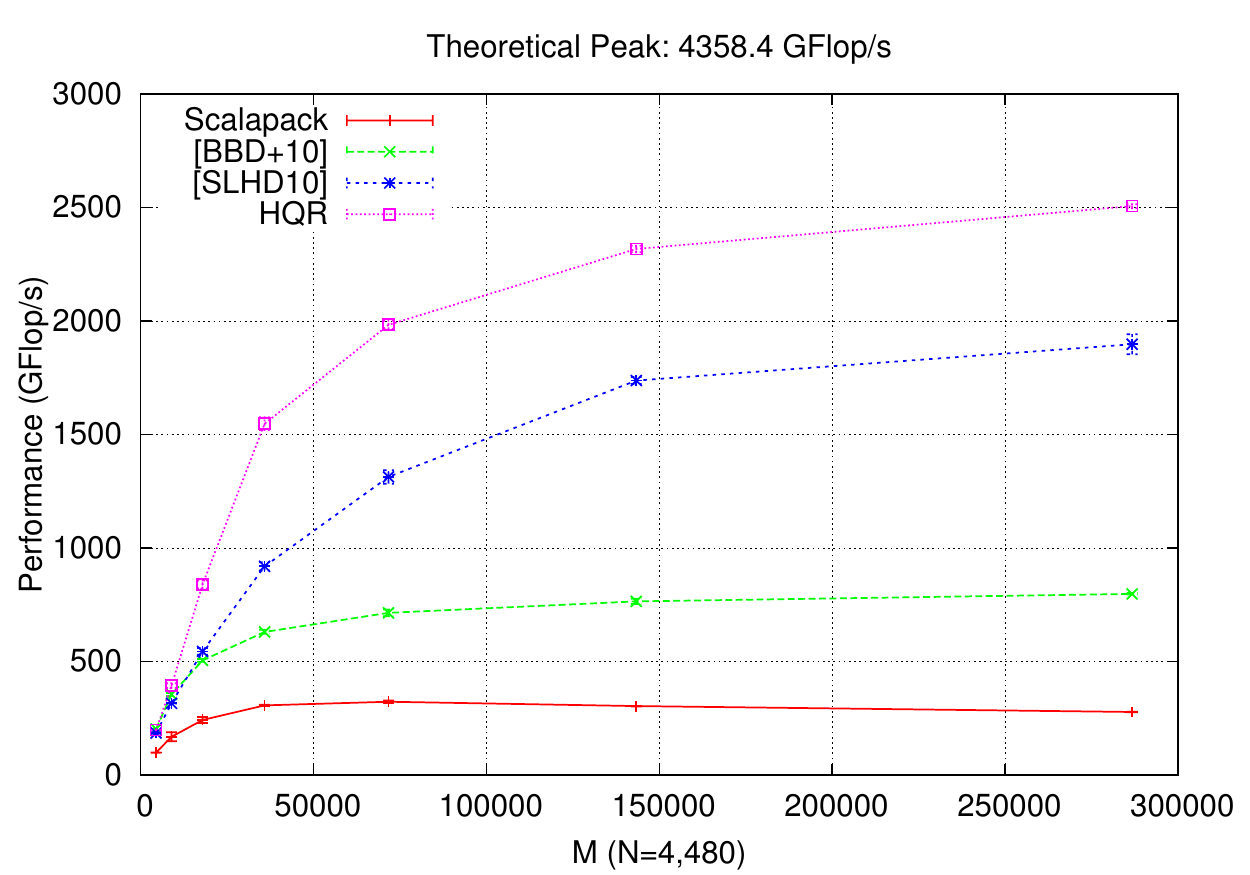}
\caption{Comparison of performance for different algorithms, on a
  $M\times 4,480$ matrix.\label{fig:comp:m}}
\end{center}
\end{figure}

\begin{figure}
\begin{center}
\includegraphics[width=.9\linewidth]{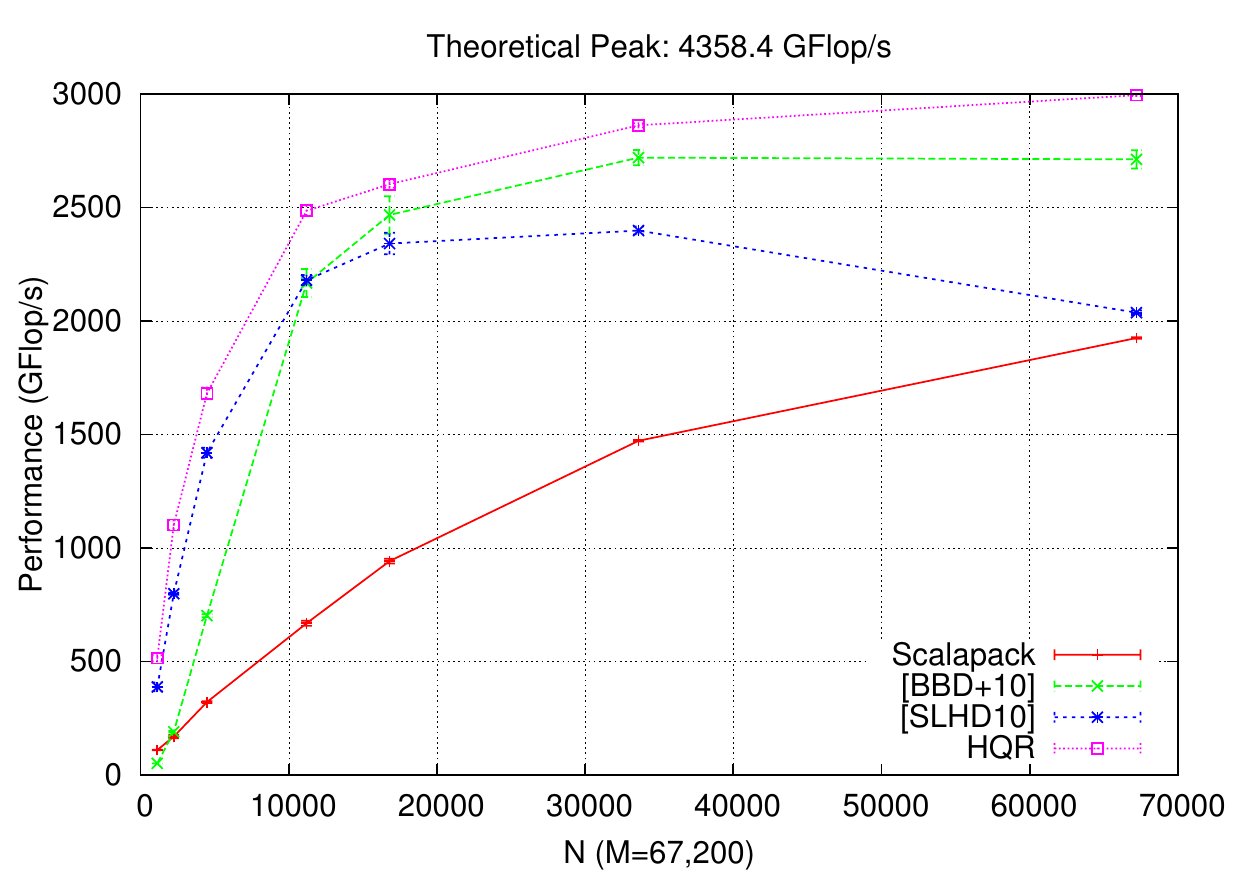}
\caption{Comparison of performance for different algorithms, on a
  $67,200\times N$ matrix.\label{fig:comp:n}}
\end{center}
\end{figure}

Figures~\ref{fig:comp:m} and~\ref{fig:comp:n} compare the performance
of the \Dague implementation of the \hqr algorithm with the
\Dague implementation of [BBD+10] and [SLHD10],
 and with the MKL implementation of the \scalapack
algorithm.

{\em $N$ fixed, $M$ varies from square to tall and skinny.}
 In Figure~\ref{fig:comp:m}, we evaluate the performance on
various matrices, from a square $16\times 16$ tiled matrix to a tall
and skinny $1,024\times 16$ tiled matrix. 
This is the same matrix set as Figure~\ref{fig:a-histo} and Figure~\ref{fig:domino-histo}.
We need low and high level trees
adapted for tall and skinny matrices so we set both level trees to \MC.
The TS level trades off some parallelism in the intra-level reduction to enable the use of TS sequential kernels which are more efficient than the TT sequential kernels.
Since, in this experiment, the local matrices have a large number of rows with respect to the number of cores on the node,
there is enough intra-node parallelism within a column reduction 
to afford a TS level, so we set $a=4$.
 Finally in the tall and skinny case,
we really want a coupling level in order to decouple the low level
tree from the inter-processor communication, so we activate the domino
optimization.
\hqr scores 2,505 GFlop/s (57.5\% of peak).


The algorithm in \scalapack is not ``tiled'', so it is not ``communication
avoiding''. The algorithm performs one parallel distributed reduction per
column, this contrasts with a tiled algorithm which performs one parallel
distributed reduction per tile.  As a consequence, there is a factor of $b$ in
the latency term between both algorithms. For a tall and skinny matrix, the
algorithm in \scalapack is indeed not compute-bounded but latency-bounded and obtains
at best 277 GFlop/s (6.4\% of peak).


The main performance bottleneck for [BDD+10] is the use of \FT. \FT has a long
start-up time to initiate the first column, it operates sequentially on the
tiles along the first tile column so that there are as many TS kernels pipelined the one after the other
as there are tiles in a
column (that is $m$,.e.g, =1024 in the
largest example considered here).  This is not suitable when there are only
$n=16$ tile columns to amortize the pipeline startup cost.  Another issue with
[BDD+10] is that the algorithm does not take into account the 2D block cyclic
distribution of the data.  This has a secondary negative impact on the
performance.
For a tall and skinny matrix, the
algorithm in [BDD+10] suffers from a long pipeline on the first tile column.
The length of this pipeline is $m$ the number of row tiles or the whole matrix.
the algorithms scores
at best 798 GFlop/s (18.3\% of peak).

[SLHD10] has been specially designed for tall and skinny
matrices~\cite{SongLHD10}.  The negative load imbalance that occurs by using a
1D block data distribution instead of a 2D block cyclic distribution is not
significant for tall and skinny matrices. At the inter-node level, the use of \BT is a good solution.
Yet, the use of TS \FT at the intra-node level is not appropriate when the local matrices
have many rows. As in [BDD+10], a long pipeline is instantiated.
A better tree is needed at the intra-node level.
For a tall and skinny matrix, the
algorithm in [SLHD10] suffers from a long pipeline on the first tile column.
The length of this pipeline is $m/p$, the number of row tiles held by a node
(which is an improvement with respect to [BDD+10] but yet too much). The algorithms scores
at best 1,897 GFlop/s (43.5\% of peak).

{\em $M$ fixed, $N$ varies from tall-skinny to square.}
In Figure~\ref{fig:comp:n}, we evaluate the performance from a tall
and skinny $240\times 4$ tiles matrix to a square $240\times 240$
tiles matrix. The high-level tree is set to \FT, while the
low-level tree is set to \MC. Depending on the value of
$N$, we choose different values for $a$: $a=1$ for small values of $N$, and
$a=4$ for larger values. Similarly, the domino coupling optimization is de-activated
once the parallelism due to the number of columns of tiles is
sufficient enough to avoid starvation, and the efficiency of the
kernels becomes more important. The choice of the \FT
high-level tree is guided by the same reason: once the parallelism is
high enough to avoid starvation, the \FT ensures a significantly smaller
number of inter-node communications.
 
[BDD+10] performs well on square matrices, however it suffers from its
more demanding communication pattern than the \hqr algorithm
(since it does not take into account the 2D block cyclic distribution of the data).
[SLHD10]
performs better on tall and skinny matrices, however the 1D data
distribution implies a load imbalance that becomes paramount when the
matrix becomes square. This is illustrated by the ratio of performance
between \hqr and [SLHD10]: on the square matrix, \hqr reaches 3TFlop/s,
while [SLHD10] reaches 2TFlop/s, thus 2/3 of the performance, as
predicted in Section~\ref{sec.related}. Likewise, when $N=M/2$,
[SLHD10] reaches 2.4TFlop/s, and \hqr 2.9TFlop/s, and $2.4/2.9 \approx
5/6$, as predicted by the model. 
Although the performance of \scalapack is lagging behind the performances of the other tile based algorithms,
\scalapack builds performance as $M$ increases and score a respectable 1,925 GFlops/sec (44.2\% of peak) on a square matrix.




\section{Conclusion}
\label{sec.conclusion}

We have presented \hqr, a hierarchical QR factorization algorithm which introduces several innovative
components to squeeze the most out of clusters of multicores. On the algorithmic side,
we have designed a fully flexible algorithm, whose many levels of tree reduction each
significantly contributes to improving state-of-the-art algorithms. A key feature is that the high level
specification of the algorithm makes it suitable to an automated implementation with the \Dague framework.
This greatly alleviates the burden of the programmer who faces the complex and concurrent programming
environments required for massively parallel distributed-memory machines.
 
 On the experimental side, our algorithm dramatically outperforms all competitors, 
 which can be seen as a major achievement 
 given (i) the ubiquity of  QR factorization in many application domains; and (ii) the vast amount of efforts that have been
 recently devoted to numerical linear algebra kernels for petascale and exascale machines.
Our implementation of the new algorithm with the \Dague
scheduling tool significantly outperforms currently available QR factorization
softwares for all matrix shapes, thereby bringing a new advance in numerical linear algebra for
petascale and exascale platforms. More specifically, our experiments on the Grid'5000 edel platform show the
following gains at both ends of the matrix shape spectrum:
\begin{compactitem}
\item On tall and skinny matrices, we reach 57.5\% of theoretical computational peak performance,
to be compared with 
6.4\% for \scalapack (9.0x speedup), 
18.3\% for [BDD+10] (3.1x),  and 43.5\% for [SLHD10] (1.3x)
\item On square matrices, we reach 68.7\% of theoretical computational peak performance,
to be compared with 
44.2\% for \scalapack (1.6x), 
62.2\% for [BDD+10] (1.1x), 
and 46.7\% for [SLHD10] (1.5x).
\end{compactitem}

Future work includes several promising directions. From a theoretical perspective, we could compute
critical paths and assess priorities to the different elimination trees. This is a very promising but technically challenging
direction, because it is not clear how to account for the different architectural costs, and because of the huge
parameter space to explore. From a more practical perspective, we could perform further experiments on machines equipped
with accelerators (such as GPUs): again, the flexibility of the \Dague software will dramatically ease the design of
\hqr on such platforms, and will enable us to explore a wide combination of reduction trees and priority settings.

\newcommand{\BIBdecl}{\setlength{\itemsep}{0.03\baselineskip}} 
\bibliographystyle{IEEEtran}
\bibliography{biblio}

\end{document}